\documentstyle[12pt]{article}

\begin{document}

\title{Perturbative QCD analyses of the nonleptonic $B$ meson decays}
\author{Ciprian Dariescu\thanks{MONBUSHO fellow, on leave of absence 
from The Dept. of Theoretical Physics,
 ^^ ^^ Al.I. Cuza'' University, 6600 Iasi, Romania} 
$^{\ddagger}$  and  Marina-Aura Dariescu\thanks{
On sabbatical leave of absence from the Dept. of Theoretical 
Physics, ^^ ^^ Al.I. Cuza'' University, 6600 Iasi, 
Romania} \thanks{E-mail: marina@jodo.sci.toyama-u.ac.jp ,
Fax: 81-764-41-2972 (before March 1, 1996).} \\ 
{\it Department of Physics, Toyama University} \\ 
{\it Gofuku, 930 Toyama, Japan}} 
\date{}
\maketitle

\begin{abstract}
Based on Brodsky-Lepage approach, the nonleptonic $B$ meson decay branching ratio is derived in terms of three parameters and is geometrically analysed as the 3-surface embedded in the 4-dimensional (abstract) Euclidean space generated by the three parameters and $BR$. Investigating its 2-dimensional sections, we find ranges for these parameters imposed by a comparison with experimental data.  
\end{abstract}

\newpage
 
{\bf 1}. Assuming the factorization, both nonleptonic and semileptonic $B$ meson decays are satisfactorily described by the heavy-flavor-symmetry approach and the results are in good agreement with observations \cite{Cle:phy}. Working in a perturbative QCD framework, Szczepaniak, Henley and Brodsky \cite{Szc:phy} have derived, in the case of nonleptonic $B$ decays, the longitudinal and transverse form factors for the {\it heavy-to-light} transitions. Since in the case of neglecting both masses in the final
states the branching ratio estimations lie much below the
experimental and other theoretical predictions upper limits,
we claim that by introducing a small mass parameter we still are
in the {\it heavy-to-light} transition approach, but raise the
numerical values of the branching ratios.
As previously, \cite{Dar:rom}, we are going to keep ourselves in
Brodsky-Lepage approach and perform a discussion on the nonleptonic decays branching ratios depending on the $B$ wave function parameter and on the final states masses and decay constants. The obtained results point out not only major differences between branching ratios corresponding to $b \rightarrow c$ and $b \rightarrow u$ transitions, but also a strong dependence on the mass parameter, even for small values of it. The same model may be applied also
in the study of rare $B$ decays, by extending the factorization to the so-called penguin diagrams \cite{Dar:plb}.
The estimations have to be compared to those belonging to Ali and Greub \cite{Ali:phy} and Ali, Ohl and Mannel \cite{Ali:plb}. 

Using the strong consequences of the heavy quark effective theory \cite{Isg:nuc} one can extract the values for the model parameters which fit this approach. As an example, the
$\bar{B}^{0} \rightarrow D^{(*)+} \pi^{-} (\rho^{-})$ decay amplitude derived in \cite{Hai:phy} will test the accuracy of the method emploied in the present work.  \\

{\bf 2}. Following \cite{Szc:phy}, we assume factorized diagrams dominated by a single gluon exchange with the spectator quark and neglect the final state interactions. For an exclusive nonleptonic decay of the $B$ meson into two much lighter $0^{-}$ or $1^{-}$ mesons, the decay amplitude is expressed as a convolution of a collinear irreducible hard-scattering quark-gluon amplitude and the following mesons wave functions \cite{Szc:phy}: \\
- For the $B$ meson of mass $M$ and decay constant $f_{B}$ we shall use the expression
\begin{equation}
\phi_{B}(x) = \varphi_{B} (x) (\gamma \cdot P_{B} + M)\gamma_{5}
\end{equation}
where the distribution amplitude \cite{Szc:phy,Dar:rom}
\begin{equation}
\varphi_{B}(x) = \frac{f_{B}}{12} \, \frac{x^{2} (1-x)^{2}}{[a^{2} x +(1-x)^{2}]^{2}} \cdot \left\lbrace \int_{0}^{1} \frac{x^{2}(1-x)^{2}}{[a^{2} x+(1-x)^{2}]^{2}} \; dx \right\rbrace^{-1}
\end{equation}
contains the parameter $a \sim 0.05-0.1$ related to the maximum of $\varphi_{B}(x)$ in the $B$ - meson; \\
- For the light $0^{-}$ meson, with $m_{1}^{2} = m^{2} << M^2$
and the decay constant $f_{1}$, the expression 
\begin{equation}
\phi_{1} = \varphi_{1}(y)(\gamma \cdot P_{1} + m) \gamma_{5},
\; \;
{\rm with} \; \;
\varphi_{1}(y) = \frac{f_{1}}{12} \, y(1-y) \left \lbrace\int_{0}^{1} y(1-y) \; dy \right \rbrace^{-1} .
\end{equation}
will bring in the estimated branching ratios the mass parameter $z=m/M$ dependence.

Working in the assumption $z^{2} \approx 0$,
we start with the unperturbed effective weak hamiltonian
\begin{equation}
H = \frac{G_{F}}{\sqrt{2}} \mid V \mid J_{\mu}^{+} J^{\mu},
\end{equation}
where $\mid V \mid$ are the quarks mixing matrix elements, and considering
\begin{equation}
J^{\mu} = f_{2} q^{\mu}, {\rm \; where \; \;} q= P_{B} - P_{1} \; 
{\rm satisfies} \; \; q^{2} = m_{2}^{2} \approx 0,
\end{equation}
we evaluate the hadronic current between the two remaining states as (see fig.1): \\
\marginpar{Figure 1}
\\
\begin{eqnarray}
{J_{\mu}} = g_{s}^{2} \left\lbrace {\rm Tr}\left[{\bar{\phi}}_{1} \gamma_{\mu } \frac{\gamma \cdot k_{b} +M}{k_{b}^{2} - M^{2}} \, \gamma_{\alpha} \, \phi_{B} \, \gamma^{\alpha} \, \frac{\lambda_{a} \lambda^{a}}{Q^{2}}\right] + \right. \nonumber \\* + \left. {\rm Tr}\left[{\bar{\phi}}_{1} \gamma_{\alpha}\frac{\gamma \cdot {k}_{1}+m}{k_{1}^{2} - m^{2}} \gamma_{\mu} \phi_{B} \, \gamma^{\alpha} \, \frac{\lambda_{a} \lambda^{a}}{Q^{2}}\right] \right\rbrace , 
\end{eqnarray}
Here, $\gamma \cdot k \; {\rm is} \; \gamma^{\beta} k_{\beta}$ while the Tr means trace over spin, flavor, color indices and integration over momentum fractions. 

Since the gluon exchanged carries the momentum
\begin{equation}
Q^{2} \approx -(1-x)(1-y) M^{2}
\end{equation}
the expression (4) comes to:
\begin{equation}
H = \frac{G_{F} g_{s}^{2} f_{B}}{9 \sqrt{2}} k I
\end{equation}
where we have used the notations:
\begin{eqnarray}
I & = & \int_{0}^{1} \frac{\varphi_{B}(x)}{1-x} dx \int_{0}^{1-a} \frac{\varphi_{1}(y)}{(1-y)^{2}} \, [y(1-2z)+(z-2)] dy + \nonumber \\* & + & \int_{0}^{1} \frac{\varphi_{B}(x)}{(1-x)^{2}} \,
z(1-2x) dx \int_{0}^{1-a} \frac{\varphi_{1}(y)}{1-y} dy \; + \; {\cal O}[z^{2}]
\end{eqnarray}
for the integration over momentum fractions and
\begin{equation}
k = \mid V \mid f_{1} f_{2}
\end{equation}
for the process-dependent constant.

We proceed now to the determination of the branching ratio for the {\it heavy-to-light} nonleptonic two body decay, namely
\begin{equation}
BR(B \rightarrow L_{1} + L_{2}) = {\Gamma(B \rightarrow L_{1}+L_{2})}/{\Gamma_{tot}}.
\end{equation}
Using the calculation of $\Gamma_{tot}$ given in \cite{Ali:phy}
and assuming
\begin{equation}
g_{s}^{2} \approx 4.77 , \; M \approx 5280 {\rm MeV} , \; f_{B} \approx 200 {\rm MeV} 
\end{equation}
we find the expression
\begin{equation}
BR = 7.3 \, \cdot \, 10^{-15} k^{2} I^{2} .
\end{equation}
which, for the particular case when $z \approx 0$, simplifies to
\begin{eqnarray}
 BR & = & 7.3 \, \cdot \, 10^{-15} k^{2} I'^{\, 2},  \nonumber \\* 
{\rm with} \; I' & = & \int_{0}^{1} \frac{\varphi_{B}(x)}{1-x} dx \int_{0}^{1-a} \frac{\varphi_{1}(y)}{(1-y)^{2}} \, (y-2) dy  
\end{eqnarray}
\\ 

{\bf 3}. Next, we shall analyse how the perturbative QCD calculations described above are satisfactorily in studying the heavy mesons decays and find the range for $a$ to account for the experimental data within the Szczepaniak, Henley and Brodsky's framework. Therefore, let us consider the $BR$ of the nonleptonic
$B$ - meson decay given by (13) as being dependent of the mass-parameter $z$, the decay parameter $k$ and the $B$ wave function parameter $a$ (see figure 2). \\
\marginpar{Figure 2}
\\
As stressed in \cite{Szc:phy}, the domain of interest corresponds to $a \in [0.05, 0.1]$. For smaller $a$'s, the $BR$ ratio increses rapidly to non-physically values while for a much bigger $a$, for example $a=0.5$ corresponding to
$< k_{\perp}^{2} >/ M^{2} \approx 0.25$,
the $BR$ is less than $5 \, \cdot \, 10^{-6}$.
So, we may conclude that out of this range of $a$, the  model fails. A comparison of the two surfaces corresponding to
$a=0.05$ and $a = 0.1$ points out small stable $BR$ values
on the $a = 0.1$ surface, smoothly increasing to $BR \approx 10^{-3}$ for $k \approx 1600$ and $z$ around 0.3.
This $k$ value can be reached in $B^{+} \rightarrow \bar{D}^{0} + D_{s}^{+}$ decay but, in order to fit the experimental $BR$ \cite{Cle:phy}, we should translate to smaller $a$ values.
On the contrary, when $a=0.05$ and $k$ is around 1600 the $BR$ is rapidly increasing from $10^{-3}$ (for $z \approx 0$) to
$1.3 \, \cdot \, 10^{-2}$ (for $z \rightarrow 0.35$).
So, once the mass of the final state increases in comparison to $M$, its contribution to the $BR$ at large values of $k$ becomes dominant as $a$ goes to 0.05. Using the experimental input on different branching ratios one is able now to determine the corresponding range for the parameter $a$. For example, in the case of the nonleptonic decay $B^{-} \rightarrow D^{0} + \pi^{-}$,
where $k = 936$ and $z \approx 0.35$, a $BR$ of about $4.5 \, \cdot \, 10^{-3}$ corresponds to the lowest limit $a = 0.05$.

An intriguing feature is to check the model for the $B \rightarrow V^{*} + L$ decays, where $V^{*}$ is an $1^{-}$ light meson, by changing (3) into
\begin{equation}
\phi_{1} = \varphi_{1}(y) (\gamma \cdot P_{1} + m) (\gamma \cdot \varepsilon),
\end{equation}
which includes the polarization of the vector meson. \\
\marginpar{Figure 3}
\\
In this case, one gets the $z$ - independent branching ratios (14), 
which leads to very small values for the $BR$ (see figure 3). 
As previously, the small $k$ transitions, like the suppressed $b \rightarrow u$, have practically an $a$ -independent $BR$, while the $b \rightarrow c$ transitions (possessing large values of $k$)
exhibit smoothly $a$ - dependent branching ratios, with an upper value around $10^{-3}$, for $a \approx 0.05$ and $k \approx 1600$. \\
\marginpar{Figure 4}
\\
Now, by comparing (see figure 4) the $BR(B \rightarrow X + Y)$ surface (as a function of $a$ and $k$ for an arbitrary $z$)
to the $z$ - independent $BR(X^{*} + Y)$ (plotted in figure 3)., we conclude that the well known HQET result \cite{Isg:nuc}
\begin{equation}
\Gamma(B \rightarrow X + Y) = \Gamma(B \rightarrow X^{*} + Y)
\end{equation}
can be accomodated in this approach for $z=0$.
Although the shape of graphs is  the same, the $z$ - dependence of $BR(B \rightarrow X+Y)$ shifts the regions with
large $k$'s to bigger branching ratio values.
Even for small values of $z$, such as $z=0.1$, the $BR(B \rightarrow X + Y)$ gets bigger than the one corresponding to $B \rightarrow X^{*} + Y$, especially when $a$ goes
close to the lower limit.
Clearly, as soon as we set $z=0$ in (9) the two surfaces coincide. \\  

{\bf 4}. As a final example, let us consider the process
$B_{d}^{0} \to D^{-} \pi^{+}$ and make an estimation of its
branching ratio and of $CP$ asymmetry parameter.
Starting with the Hamiltonian (4) and neglecting the exchange diagrams, we obtain for the nonleptonic $B_{d}^{0}$ decay into non-$CP$ eigenstates $D^{-} , \pi^{+}$ the amplitudes:
\begin{eqnarray}
& a) & \langle D^{-} \pi^{+} \mid H \mid B_{d}^{0} (0) \rangle \approx \frac{G_{F}}{\sqrt{2}} V_{cb}^{*} V_{ud} f_{\pi} M^{2} F^{BD} (0) \nonumber \\* & \nonumber & \\*
& b) & \langle D^{-} \pi^{+} \mid H \mid {\bar{B}}_{d}^{0} (0) \rangle \approx \frac{G_{F}}{\sqrt{2}} V_{ub} V_{cd}^{*} f_{D} M^{2} F^{B \pi} (0),
\end{eqnarray}
and their complex conjugated corresponding to 
$\langle D^{+} \pi^{-} \mid H \mid {\bar{B}}_{d,phys}^{0} (t) \rangle$ and 
$\langle D^{+} \pi^{-} \mid H \mid B_{d,phys}^{0} (t) \rangle$.
The model dependent form factors:
\begin{eqnarray}
F^{BD}(0) & = & \frac{g_{s}^{2} f_{B} f_{D}}{8 M^{2}} \left \lbrace
\int_{0}^{1} \frac{\varphi_{B}(x)}{1-x} dx \int_{0}^{1-a} \frac{\varphi_{D}(y)}{(1-y)^{2}} \, [y(1-2z)+z-2] \,
dy + \right. \nonumber \\* 
& + & \left. \int_{0}^{1-a} \frac{\varphi_{D}(y)}{1-y} dy \int_{0}^{1} \frac{\varphi_{B}(x)}{(1-x)^{2}} \, z(1-2x) \, dx \right \rbrace , \; {\rm where} \; z=\frac{m_{D}}{M} \approx 0.2 \nonumber \\*
F^{B \pi}(0) & = & \frac{g_{s}^{2} f_{B} f_{\pi}}{8 M^{2}} \int_{0}^{1} \frac{\varphi_{B}(x)}{1-x} dx \int_{0}^{1-a} \frac{\varphi_{\pi}(y)}{(1-y)^{2}} \,  (y-2) \, dy 
\end{eqnarray}
lead to (13) and (14) - type branching ratios respectively.
In the first case, $k = 936$ and $z \approx 0.35$,
the numerical values of branching ratio are in the range $4.4 \, \cdot \, 10^{-3}$ (for $a = 0.05$) and $4.7 \, \cdot \, 10^{-4}$ (for $a=0.1$) (see figure 2). A comparison to the experimental
data $BR(B^{0} \to D^{-} \pi^{+}) = (3.0 \pm 0.4) \cdot 10^{-3}$
set $a$ in the range [0.054 - 0.059].
The second case is much supressed because of $V_{ub}$ and corresponds to $k = 34$ and $z = 0$ (see figure 3), the numerical values of $BR$ being around $10^{-7}$.
Following the procedure described in \cite{Don:Phy}, we compute the $CP$ asymmetry parameter
\begin{equation}
{\cal A}_{D^{-} \pi^{+}} = \frac{\mid \langle D^{-} \pi^{+} \mid H \mid B_{d,phys}^{0}(t) \rangle \mid^{2} - \mid \langle D^{+} \pi^{-} \mid H \mid {\bar{B}}_{d,phys}^{0}(t) \rangle \mid^{2}}
{\mid \langle D^{-} \pi^{+} \mid H \mid B_{d,phys}^{0}(t) \rangle \mid^{2} + \mid \langle D^{+} \pi^{-} \mid H \mid {\bar{B}}_{d,phys}^{0}(t) \rangle \mid^{2}},
\end{equation}
whose numerical values, after performing the usual
time integration, are between $-2.41 \, \cdot \, 10^{-3}$
(for $a=0.05$) and $-3.24 \, \cdot \, 10^{-3}$ (for $a=0.1$). \\

{\bf 5}. In conclusions, working in the approach of Szczepaniak, Henley and Brodsky \cite{Szc:phy} we have derived the branching ratio corresponding to the nonleptonic {\it heavy-to-light}
$B$ meson decays in terms of three parameters:
the $a$ parameter related to the momentum distribution in the wave function of the $B$ meson, the mass parameter $z$
whose second power in the form factor is neglected,
and the $k$ parameter containing the decay constants of the
mesons in the final state and the CKM matrix elements.
We have performed a geometrical analyses of the
$BR(a, k, z)$ as a 3-surface and of its 2-dimensional sections corresponding to $a=const.$ or $z=const.$ and imposed ranges
for the parameters in order to ensure an agreement with the observations. For a given process, 
as for example $B_{d}^{0} \to D^{-} \pi^{+}$,
since $z$ and $k$ are known, the comparison to experimental data fixes the value of $a$ in the range [0.054, 0.059].
In this particular case,
the computed values of the $CP$ asymmetry parameter
are close to the one obtained in the Bauer-Stech-Wirbel scheme \cite{Bau:ZPh}, namely $-5 \, \cdot \, 10^{-3}$ \cite{Don:Phy},
the largest difference being of a factor 2, for $a=0.05$. \\ 

The authors are greatly indebted to Professor E.A. Paschos
for reading the manuscript with a friendly disposition
and providing valuable insights and suggestions which helped in
improving its original version.
It is a pleasure to stress the fertile environment of Toyama
University and especially the kind hospitality of the Quantum Theory Group from the Physics Department while this work was being done. 
(C.D.) expresses his gratitude to the Japanese Government for
financially supporting his work under a MONBUSHO Fellowship.
(M.A.D.) thanks Professor D. Wyler and Dr. H. Simma for
having stimulated her interest in Brodsky-Lepage approach.

\newpage

\begin{center}
The list of figure captions
\end{center}
Caption 1. The contributing diagrams in the current matrix element $J_{\mu}$ \\ 
Caption 2. $BR(B \rightarrow L_{1} + L_{2})$ for $a \in [0.05-0.1]$. \\
Caption 3. The $\lbrace a,k \rbrace$ - dependence of the $BR(B \rightarrow X^{*} + Y)$. 
\\
Caption 4. $BR(B \rightarrow X + Y)$ for $z=0.1$ (the upper sheet) compared to the $z$ - independent $BR(B \rightarrow X^{*} + Y)$

\begin{figure}[p]
\vspace*{10cm}
\caption{The contributing diagrams in the current matrix element $J_{\mu}$}
\end{figure}
\begin{figure}[p]
\vspace*{12cm}
\caption{$BR(B \rightarrow L_{1} + L_{2})$ for $a \in [0.05-0.1]$.}
\end{figure}
\begin{figure}[p]
\vspace*{12cm}
\caption{The $\lbrace a,k \rbrace$ - dependence of the $BR(B \rightarrow X^{*} + Y)$.}
\end{figure}
\begin{figure}
\vspace*{12cm}
\caption{$BR(B \rightarrow X + Y)$ for $z=0.1$ (the upper sheet) compared to the $z$ - independent $BR(B \rightarrow X^{*} + Y)$.}
\end{figure}

\end{document}